\documentclass[manuscript]{acmart}

\AtBeginDocument{%
  }

\setcopyright{acmlicensed}
\copyrightyear{2018}
\acmYear{2024}
\acmDOI{}

\acmConference[CHI '25]{In Proceedings of the CHI Conference on Human Factors in Computing Systems}{}{}
\acmISBN{978-1-4503-XXXX-X/18/06}

\usepackage{graphicx}
\usepackage{subcaption}
\usepackage{booktabs}
\usepackage{xcolor}
\usepackage{colortbl}
\definecolor{myblue}{rgb}{0.94,0.97,1.0}

\begin{document}

\title{Evaluating the Impact of Warning Modalities and False Alarms in Pedestrian Crossing Alert System}




\author{Hesham Alyamani}
\email{halyamani@wisc.edu}
\author{Yucheng Yang}
\email{yang552@wisc.ed}
\author{David Noyce}
\email{danoyce@wisc.edu}
\author{Madhav Chitturi}
\email{mchitturi@wisc.edu}
\author{Kassem Fawaz}
\email{kfawaz@wisc.edu}
\affiliation{
\institution{University of Wisconsin-Madison}
\city{Madison}
\state{Wisconsin}
\country{USA}
}

\renewcommand{\shortauthors}{Alyamani et al.}


\begin{abstract}

With the steadily increasing pedestrian fatalities, pedestrian safety is a growing concern, especially in urban environments. Advanced Driver Assistance Systems (ADAS) have been developed to mitigate road user risks by predicting potential pedestrian crossings and issuing timely driver alerts. However, there is limited understanding of how drivers respond to different modalities of alerts, particularly in the presence of false alarms. In this study, we utilized a full-scale driving simulator to compare the effectiveness of different alert modalities, audio-visual (AV), visual-tactile (VT), and audio-visual-tactile (AVT), in alerting drivers to various pedestrian jaywalking events. Our findings reveal that, compared to no alerts, multimodal alerts significantly increased the number of vehicles stopped for pedestrians and the distance to pedestrians when stopped. However, the false alarms negatively impacted driver trust, with some drivers exhibiting excessive caution, alert fatigue and anxiety, even including one instance where a driver fully stopped when no pedestrian was present.

\end{abstract}

\maketitle

\section{Introduction}
Pedestrians, being the most vulnerable among road users, face high risks of vehicular traffic, which demands proper safety measures~\cite{paul2016}. 
Pedestrian crashes account for a significant portion of the total traffic-related injury and fatality globally~\cite{chrysler2015}. The number of pedestrians killed in traffic crashes have been steadily increasing since 2010, reaching a high of 7,522 in 2022, according to the National Highway Traffic Safety Administration~\cite{nhtsa_pedestrian_safety}. Nearly 75\% of pedestrian fatalities occur at non-intersection locations. Over 75\% of pedestrian fatalities occur in dawn, dusk, or night conditions~\cite{nhtsa_bicycle_safety, nhtsa_pedestrian_safety_countermeasures}. 
At the urban level, the interaction between vehicles and pedestrians is often frequent and intricate, making a timely and accurate driver responses critical~\cite{martinez2012, habibovic2011}. 
Advanced Driver Assistance Systems (ADAS) including pedestrian crossing alerts for drivers mitigate such hazards~\cite{kuo2016}. These systems use various alerting modalities that include auditory, visual, and tactile to warn drivers and increase driver attention and responsiveness. ADAS provide additional cues to the driver in a timely pedestrian crossing alert, lowering the chances of a crash~\cite{gu2022}. 


Accurate alert systems are critical for providing reliable alerts about potential hazards, minimizing false alarms, and improving driver trust. However, false alarms, or false positives, occur when the alert system mistakenly signals a potential hazard. These false alarms can negatively affect driver behavior by causing distraction or leading to habituation, where drivers become less responsive to alerts over time. Frequent false alarms erode trust in the system~\cite{pang2022}, leading drivers to ignore alerts or react less urgently to genuine warnings, thereby increasing crash risks~\cite{weibull2023}. False alarms can result from sensor inaccuracies, adverse environmental conditions, and system malfunctions. Common contributors include poor sensor calibration and unfavorable weather conditions, such as fog, rain or low-light environments~\cite{anik2021}.

To explore the most effective methods for alerting drivers and understand the impact of false alarms, we conducted a between-subjects experimental study with 48 participants (14 female and 34 male), aged between 27 and 67 years old, with varying levels of driving experience. Participants operated a full-scale driving simulator to drive in a Unity-based, town-scale simulated urban environment, and we measured their responses across different road events and alert modalities. We interleaved true warnings with false alarms to evaluate the participants' responses to the driver alert system. To provide a baseline for comparison, we included scenarios with no alerts during driving.

Our contributions can be summarized as follows:
\begin{itemize}
    \item We conducted a full-scale driving simulator study that provides an in-car driving experience similar to real-world conditions, ensuring fidelity and participant safety.
    \item We designed and implemented a simulated urban driving environment using Unity, incorporating various road crossing events. The driving sessions included multiple jaywalking warnings, dummy events, and false alarms to encourage participants to interact more with the driving sessions rather than only reacting to warnings.
    \item Through participant recruitment and a post-experiment survey, we quantified the impact of different modalities for pedestrian crossing alert systems and gained insights into how false alarms contribute to distrust and confusion among participants. 
\end{itemize}



\section{Related Work}

\subsection{Pedestrian Vehicle Interaction}
Research on pedestrian and vehicle interaction has focused on two key areas: the development of external Human-Machine Interfaces (eHMI) for vehicles and the prediction of the intention of pedestrian crossing through cameras or LiDAR sensors on road-side infrastructure and vehicles. 

The first area of research involves the design of eHMIs, which are external interfaces installed on autonomous vehicles (AVs) to facilitate communication between the vehicle and pedestrians~\cite{mahadevan2018communicating, lanzer2023interaction, tran2024exploring, chang2024must}. These interfaces aim to enhance safety by providing alerts and indicating the vehicle's intent and awareness to pedestrians, thereby creating a safer interaction environment between pedestrians and AVs. Despite these advancements, the widespread adoption of autonomous vehicles remains in the future, and pedestrian safety continues to be a critical concern in road user safety.

Researchers also focus on predicting pedestrian crossing intent, which is crucial for enhancing road safety, particularly in the context of autonomous driving and traffic management~\cite{zhang2023pedestrian}. Current methods for predicting pedestrian crossing intent primarily rely on visual information obtained from cameras or LiDAR sensors mounted on vehicles~\cite{yang2022predicting} or roadside infrastructure, such as CCTV cameras installed at intersections~\cite{zhang2021pedestrian}. These methods analyze captured images to identify various aspects of the pedestrian's current state, including their motion~\cite{volz2018inferring, volz2016data}, past trajectory~\cite{zhao2019probabilistic}, skeletal structure~\cite{zhang2021pedestrian, fang2018pedestrian, piccoli2020fussi, lorenzo2020rnn}, and even demographic factors such as age and gender~\cite{zhang2020research}. By applying deep learning techniques to these data~\cite{zhang2023pedestrian}, researchers aim to accurately determine whether a pedestrian intends to cross the street, which is vital for preventing crashes and ensuring a safer road environment.

\subsection{Development of ADAS and Warning Modalities}

ADAS have evolved to include a range of warning systems that alert drivers to potential dangers. Early systems relied primarily on auditory alerts, using sounds to capture attention effectively~\cite{shahab2010}. While these are effective in cutting through visual distractions, their impact can be diminished by environmental noise or driver habituation~\cite{bella2021}. Visual alerts, delivered through dashboard indicators or head-up displays (HUDs), offer clear, actionable warnings without overwhelming the driver. Tactile alerts, such as vibrations in the steering wheel or seat, provide an additional layer of feedback, particularly useful when visual or auditory channels are overloaded~\cite{fan2023}.

\subsubsection{Types of Driver Alerts}

Alerting systems play a crucial role in enhancing driver awareness and ensuring timely responses.

\begin{itemize}
    \item \textbf{Auditory Alerts:} These use sound, such as beeps or alarms, to warn drivers of hazards, with their effectiveness depending on factors like sound frequency, amplitude, and duration~\cite{glatz2015}. Properly designed auditory alerts improve safety by enhancing driver focus and response~\cite{landry2019, ma2024}.

    \item \textbf{Visual Alerts:} Visual alerts, often displayed on dashboards or HUDs, provide critical cues about potential dangers. High-contrast colors and strategic placement in the driver’s line of sight enhance their effectiveness~\cite{szagala2021}. Augmented Reality HUDs further improve pedestrian awareness without adding distraction~\cite{kim2022}.
    
    \item \textbf{Tactile Alerts:} These alerts, delivered through vibrations in the steering wheel, seat, or pedals, offer a non-visual, non-auditory method of communication. They are particularly effective in noisy or visually cluttered environments, ensuring crucial alerts are noticed~\cite{dicampli2020, jeong2023}.
    
    \item \textbf{Multi-Modal Alert Systems:} Modern ADAS increasingly incorporate multi-modal systems that combine auditory, visual, and tactile alerts. This redundancy increases the likelihood that drivers will notice and respond to warnings, making these systems highly adaptable to various driving conditions~\cite{dong2023, ma2024}.
    
\end{itemize}

\subsection{Driving Simulators}
Recent advancements in driving simulators have improved virtual environments for studying driver behavior and road interactions. Goedicke et al.\cite{goedicke2022xr} developed a mixed reality simulator where participants drive a real car while wearing a VR headset, interacting with virtual objects integrated into the real world. Bu et al.\cite{bu2024portobello} extended traditional in-lab simulators to an on-road platform, enabling comparative studies between in-lab and real-world driving. Hou et al.\cite{hou2020autonomous} used VR to explore interactions between autonomous vehicles and cyclists, while Tateyama et al.\cite{tateyama2010observation} created a 180-degree simulator to study turning behaviors. Our simulator enhances the immersive experience with side and rear view mirrors and a Unity-based simulated environment. 

\section{Experiment Methods}

Our study investigates two main research questions: (1) \textit{What warning modality is most effective in alerting drivers to pedestrian jaywalking?} (2) \textit{How do drivers react, both behaviorally and emotionally, to false alarms within pedestrian crossing alert systems?}

We conducted a two-phase experiment to address the above research questions. In the first phase, the participants completed driving sessions in the driving simulator with different warning modalities. In the second phase, they provided feedback on the warning system by completing a post-session questionnaire. Our study was exempted by the university's Institutional Review Board (IRB).

\subsection{Recruitment Procedure}

Before participating in the driving simulator study, each participant read and signed a consent form, indicating their willingness to participate in the study. Next, they were introduced to the experiment details through a pre-recorded video, which provided the necessary instructions on how to follow the designated path and guidance on stopping the simulation in case of motion sickness or discomfort during driving.

Participants were not informed about the pedestrian crossing alert system or specific road scenarios prior to the experiments. They were simply instructed to follow the turn signs on the roads and drive as they would in a real-world environment. The participants had complete control over the vehicle, including speed control, avoiding other road users, and deciding when to stop.

\subsubsection{Participant Distributions}
We recruited participants through our institute's research service and social networks, ensuring a diverse demographic. 48 participants participated in the study, with 14 females and 34 males. Their ages ranged from 27 to 67 years as shown in Table~\ref{tab:age_distribution}.  All participants held a valid driver's license and were over 18 years old. 
One driving session took approximately 30 minutes to complete, and we compensated all participants with a \$20 gift card for their participation.

\begin{table}[h]
    \centering
    \begin{tabular}{ c c c c c c}
        \toprule
        \textbf{Age Range} & 25-34 & 35-44 & 45-54 & 55-64 & 65-74 \\
        \midrule
        \textbf{Number of Participants} & 25 & 20 & 1 & 1 & 1 \\
        \bottomrule
    \end{tabular}
    \caption{Age distribution}
    \label{tab:age_distribution}
\end{table}


\subsection{Experiment Setup}

\begin{figure}[h]
    \centering
    \begin{minipage}{0.49\textwidth}
        \centering
        \includegraphics[width=\textwidth]{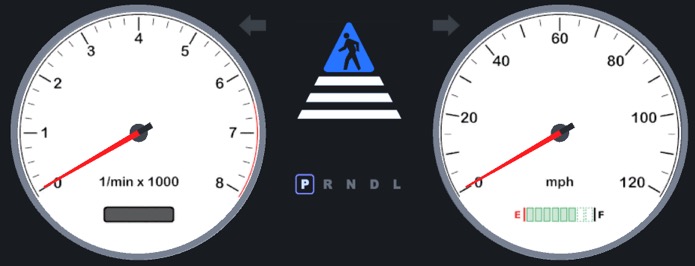}
        \caption{Visual warning}
        \label{fig:visual_warning}
    \end{minipage}
    \hfill
    \begin{minipage}{0.4\textwidth}
        \centering
        \includegraphics[width=\textwidth]{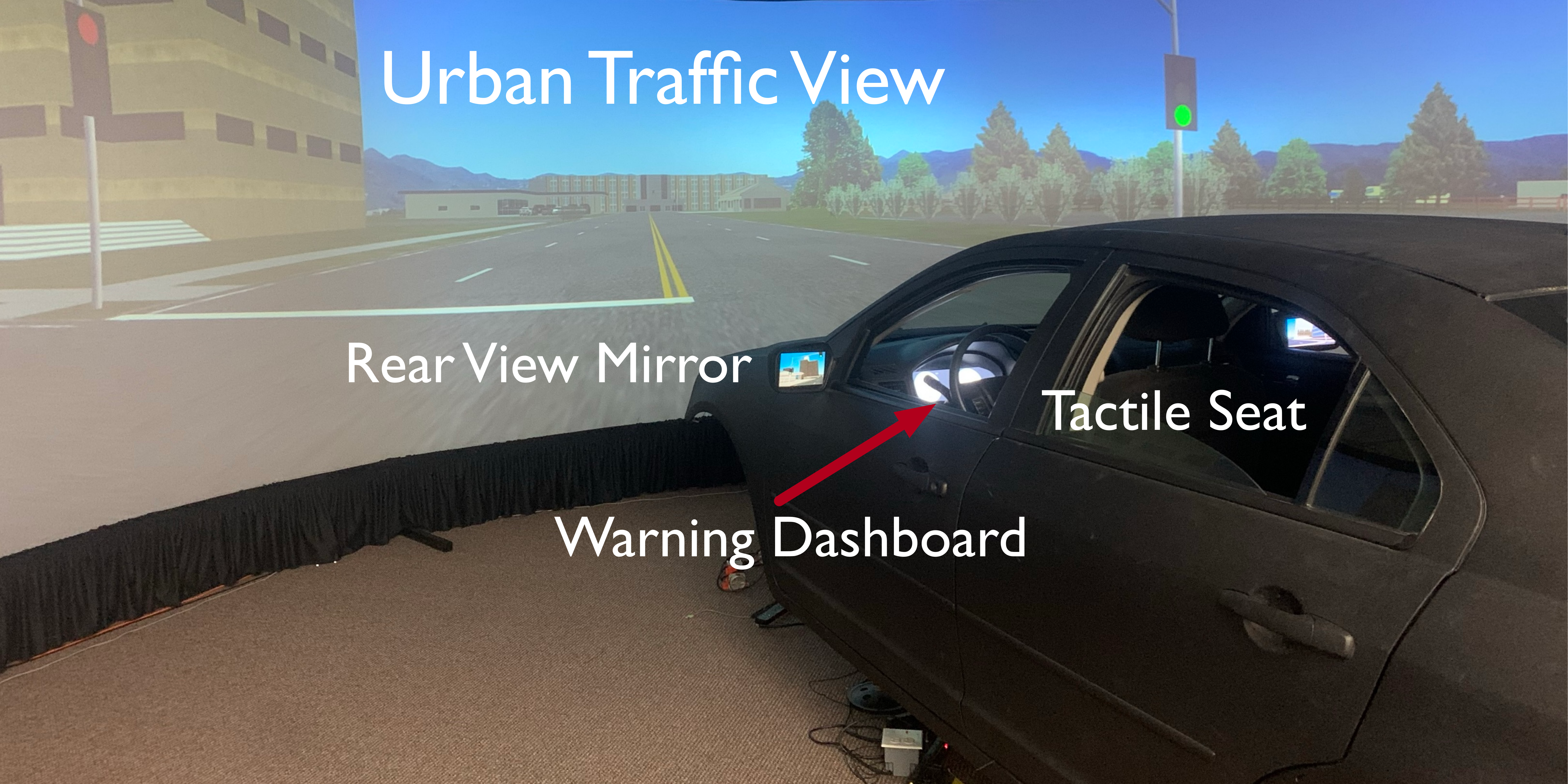}
        \caption{Full-scale Driving simulator}
        \label{fig:simulator}
    \end{minipage}
\end{figure}

Our experiment involves four driving sessions equipped with four warning modalities. These warning modalities are 1) a base scenario  condition without any kind of pedestrian crossing alerts, 2) an audio-visual (AV) warning with beeping sounds and dashboard displays as shown in Figure~\ref{fig:visual_warning}, 3) a visual-tactile (VT) warning with dashboard displays and seat vibration as in Figure~\ref{fig:simulator}, and 4) an audio-visual-tactile (AVT) warning combining the three modalities.
 
To avoid learning and carryover effects, each participant is exposed to \textbf{only one} of the four driving sessions, and therefore each driving session has 12 participants.
All four driving sessions share the same sequence of road events described in Section~\ref{sec:ped_cross_events}. 
In the sessions that included warnings, we introduced pedestrian crossing alerts for both actual crossing events and false alarms to assess drivers’ reactions.
The experiment was setup in an urban driving conditions with varying traffic volumes and pedestrian activity. 



\subsubsection{Full-Scale Driving Simulator}

We conducted the experiment in a full-scale driving simulator (shown in Figure~\ref{fig:simulator}) to study driver behavior and safety within realistic traffic environments. This simulator is based on a standard Ford Fusion, equipped with real driving interfaces for operation and various sensors to monitor essential driving input. This simulator offers a virtual platform that mimics real-world driving conditions, allowing researchers to explore how drivers respond to various traffic scenarios in a controlled and safe environment. It plays a crucial role in understanding driver performance and safety, especially in simulating both daily driving environment and high-risk driving situations.


\paragraph{Data Collection}

Operating at a 60-Hz sampling rate, the driving simulator captures key metrics such as speed, lane position, steering angle, brake pedal position, and gas pedal position. This high-frequency data collection enables an in-depth analysis of driver behavior across diverse environments, from urban intersections to highways. 



\subsubsection{Driving Session Design in Unity}
\begin{figure}
    \centering
    \includegraphics[width=0.5\linewidth]{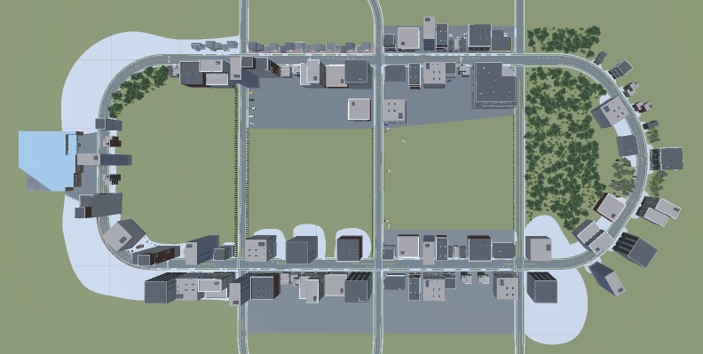}
    \caption{Overview of urban driving scenarios }
    \label{fig:overview}
\end{figure}

We developed the driving sessions with multiple pedestrian jaywalking events using the Unity 3D game engine. The bird-view in Figure~\ref{fig:overview} shows the road map for the entire scene. We designed various buildings, multiple intersections, simulated pedestrian walking, and simulated vehicles driving randomly in a highly detailed and interactive environment that mirrors the real-world driving. We included realistic pedestrian behaviors, such as walking along the street, walking behind trucks, and abrupt jaywalking using Unity's physics engine. 
Our driving session provides a robust testing ground for evaluating driver reactions toward the pedestrian crossing alert system.

\begin{figure}
    \centering
    \includegraphics[width=0.5\linewidth]{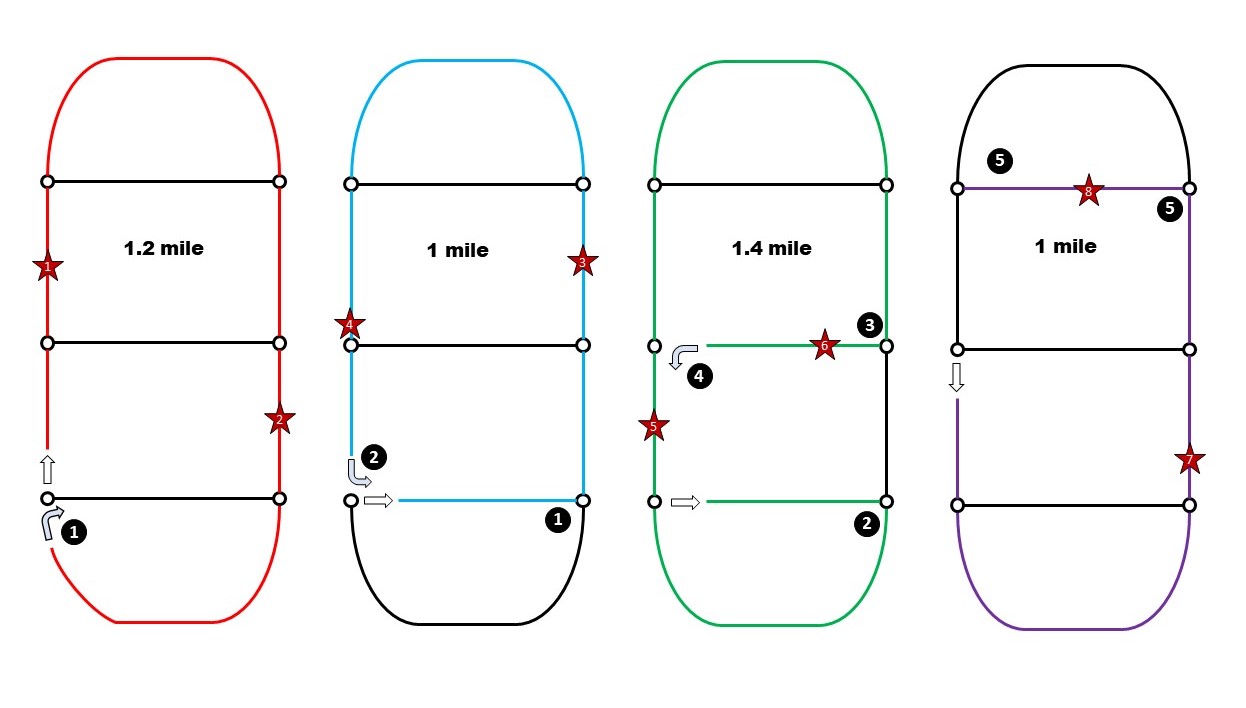}
    \caption{Run Schematic: four-loop driving route. Black lines indicate path not driven through, numbers in circles indicate start and end points, and numbers in stars represent actual pedestrian jaywalking events.}
    \label{fig:scenario_layout}
\end{figure}

As shown in Figure~\ref{fig:scenario_layout}, the driving session utilizes a closed-loop track formed by four loops connecting end to start. In other words, the end point of each loop is the start point of the next loop. The total driving route length is 4.6 miles. Each loop includes curves, tangents, and intersections to add complexity to driving environment. For each driving session, pedestrian jaywalking events are evenly distributed across four loops as described in Sec~\ref{sec:ped_cross_events}.

The simulator generates pedestrian jaywalking events based on vehicle proximity. There are two trigger points for each event: one for triggering pedestrian walking and one for triggering the pedestrian crossing. When the vehicle crosses the first trigger point, a pedestrian begins walking in the same direction as the vehicle along its future path. At this point, the participant can visually observe the pedestrian from a distance. Later, when the vehicle crosses the second trigger point, the pedestrian starts to cross, and the participant receives a pedestrian crossing alert in various modalities or no warning if in the base scenario.  



\begin{figure}[ht]
    \centering
    \begin{subfigure}[b]{0.32\linewidth}
        \centering
        \includegraphics[width=\linewidth]{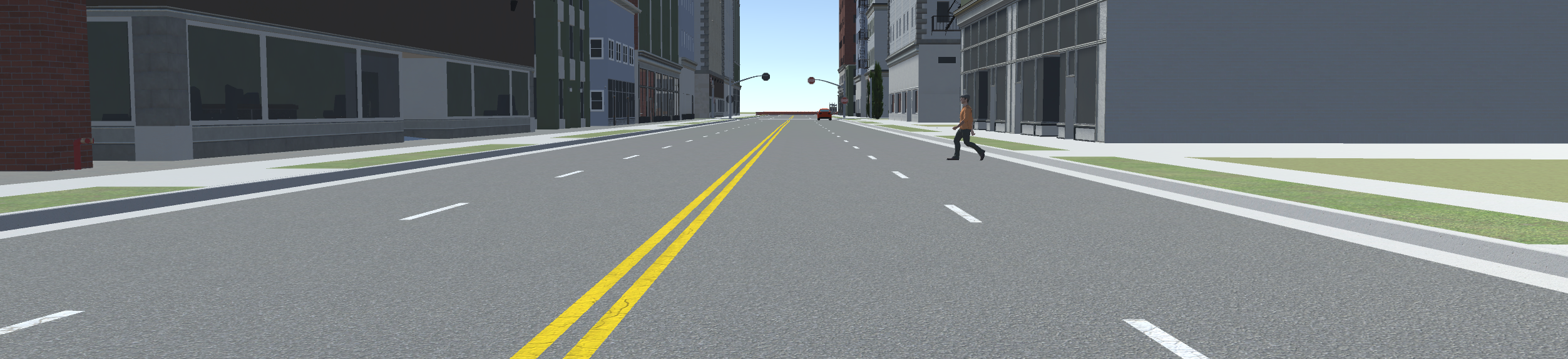}
        \caption{Right Crossing}
        \label{fig:right_crossing}
    \end{subfigure}
    \hfill
    \begin{subfigure}[b]{0.32\linewidth}
        \centering
        \includegraphics[width=\linewidth]{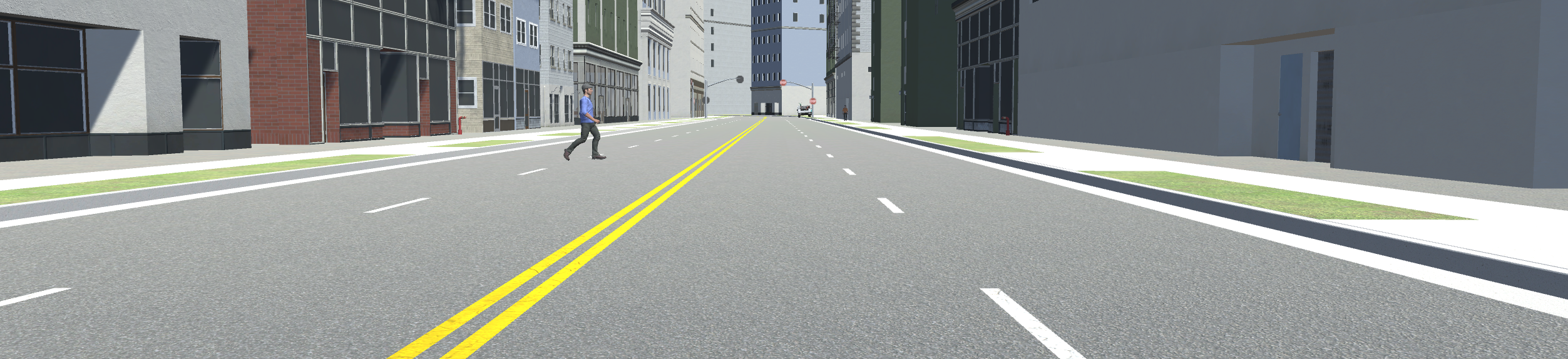}
        \caption{Left Crossing}
        \label{fig:left_crossing}
    \end{subfigure}
    \hfill
    \begin{subfigure}[b]{0.32\linewidth}
        \centering
        \includegraphics[width=\linewidth]{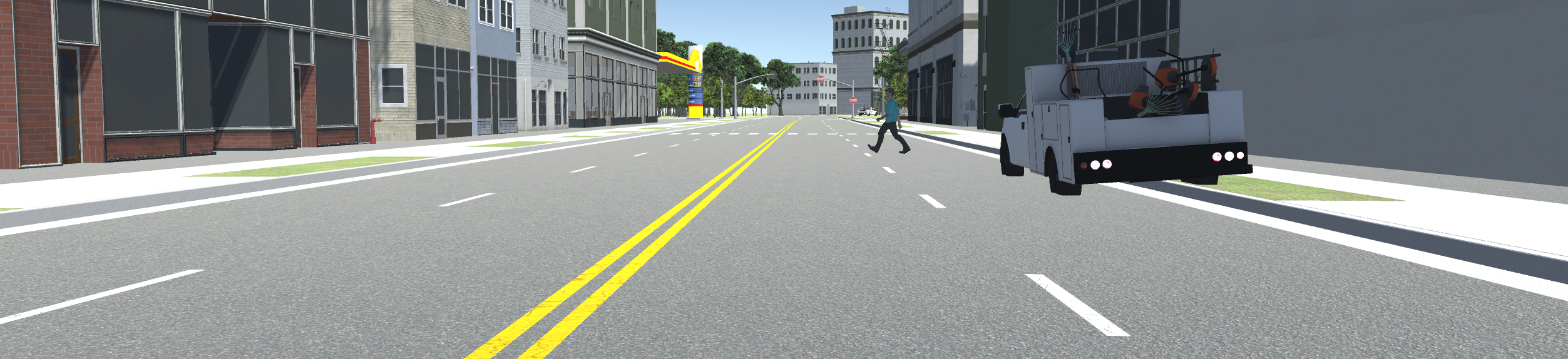}
        \caption{Truck Blocking}
        \label{fig:truck_blocking}
    \end{subfigure}

    \caption{Simulated Driving Environment Scenarios}
    \label{fig:driving_env_sample}
\end{figure}


\subsubsection{Road Events}\label{sec:ped_cross_events}

We designed different road events to test the effectiveness of pedestrian crossing alert system under various conditions.
There are four road event types with actual pedestrian jaywalking that trigger pedestrian crossing alerts: right crossing, left crossing, short distance and truck blocking. False alarm event also triggers the pedestrian crossing alert, but no pedestrian crosses. Each event type occurs twice in a driving session and differs from other types in crossing side, obstacles and warning activation distance. Details of road event types are detailed in Table~\ref{tab:events} and as follows:

\begin{itemize}
\item \textbf{Right crossing}: A pedestrian enters the road from the right side of the road, and the pedestrian crossing alert is triggered when the distance between the vehicle and the pedestrian reaches 200 feet. This serves as a basic scenario, with other crossing event type differing from it in one aspect.  

\item \textbf{Truck Blocking}: A pedestrian crosses behind a truck parked on the road side, and is not in the participant's line of sight when they begin crossing.

\item \textbf{Left Crossing}: Pedestrian enters road from the left side of the road, offering a variation on the direction of crossing. 

\item \textbf{Short Distance}: The participant receives a warning when the distance between the vehicle and the pedestrian reaches 160 feet instead of 200 feet.

\item \textbf{Dummy Event}: Dummy events involves multiple events that do no directly impact the vehicle's normal driving, such as stop sign runners, vehicles blocking the lane, or pedestrians walking along the street. Dummy events are intended to distract the participants from only focusing on crossing pedestrians. No pedestrian crossing alert will be triggered in a ``Dummy Event.''

\item \textbf{False Alarm:} No pedestrian crosses during false alarms, though the participant receives a pedestrian crossing alert. In the first false alarm, a truck is parked by the right lane. In the second, pedestrians are walking along the street. These scenarios are similar to ``Dummy Events'', but the participant receives pedestrian crossing alert in ``False Alarm''.
\end{itemize}

\begin{table}[h]
    \centering
    \begin{tabular}{l c c c c c c }
        \toprule
         \textbf{Event name} & \textbf{Number of Events} & \textbf{Crossing Side} &  \textbf{Obstacle} & \textbf{Warning Activation Distance} \\
        \midrule
         Left Crossing & 2 & Left & None & 200 feet \\
         Short Distance & 2 & Right & None & 160 feet \\
         Truck Blocking & 2 & Right & Blocked by truck & 200 feet \\
         Right Crossing & 2 & Right & None & 200 feet \\
        \bottomrule
    \end{tabular}
    \caption{Pedestrian jaywalking events for each driving session.}
    \label{tab:events}
\end{table}

\begin{figure}[h]
    \centering
    \includegraphics[width=\textwidth]{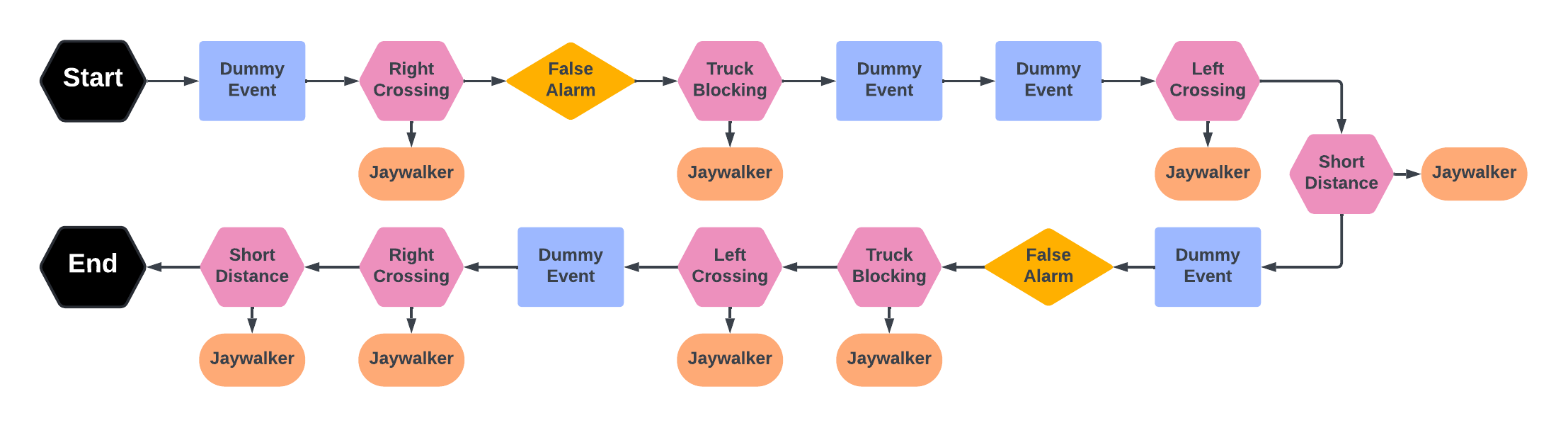}
    \caption{Sequence of events in one driving session, including six types of events, Right Crossing, Left Crossing, Truck Blocking, Short Distance, Dummy Event, and False Alarm.  Each session will utilize one of four warning modes, 1) no warning, 2) Audio-Visual (AV), 3) Visual-Tactile (VT), and 4) Audio-Visual-Tactile (AVT).}
    \label{fig:sequence_of_events}
\end{figure}

We carefully designed the sequence of road events, with eight actual crossing events, five dummy events, and two false alarms events, as illustrated in Figure~\ref{fig:sequence_of_events}.
Driving sessions with different warning modalities share the same sequence of events.
Only during the crossing events in hexagon shape, there are jaywalkers crossing the road in the middle of the block. Stop signs are placed at road intersections and they are not part of any event type. 

\subsection{Post-session Questionnaire}
After completing the driving session, participants were asked to fill out a questionnaire to assess their overall experience with the pedestrian crossing alert system, as well as their attitudes toward false alarms during driving. The survey consisted of first four multiple-choice questions and one last open-ended question, as follows:

\begin{enumerate}
    \item How clear were the system's alerts for crossing pedestrians?
    \item How timely do you find the pedestrian crossing alerts?
    \item How would you rate your overall confidence in the system for pedestrian crossing alerts?
    \item Would you recommend this pedestrian warning system to other drivers?
    \item How did the system's false alerts or failures to detect pedestrian crossing scenarios affect your driving?  (you do not need to answer if you have no alert)
\end{enumerate}

\section{Data Analysis and Findings}
We conducted both quantitative and qualitative analyses to comprehensively evaluate participants' responses to the pedestrian crossing alert system across four warning modalities.
We begin with analyzing participant reactions during the driving sessions, followed by an exploration of the participants' responses to the post-session questionnaire.
Our objective is twofold: first, to examine how the pedestrian crossing alert system supports participants in making driving decisions, and second, to assess participants' reactions to false alarms when no pedestrians are present.

\subsection{Participant Reaction in Driving Session}

Using vehicle dynamics data collected from the driving simulator, we computed the following performance indicators to assess driver behavior during the pedestrian jaywalking events and false alarms:

\begin{itemize}
    \item \textbf{Vehicle Stopping Count:} Whether vehicles fully stopped upon receiving a pedestrian crossing alert.
    \item \textbf{Brake Initiation Count:} Whether participants applied the brakes in response to a pedestrian crossing alert.
    \item \textbf{Average Stopping Distance:} The distance between the vehicle and the pedestrian when vehicle stopped.
    \item \textbf{Average Speed After Events:} The average vehicle speed after the road event and before the next event.
\end{itemize}

We analyzed these indicators across different road events in four driving sessions. Each of the four driving sessions had 12 participants, and during each session, participants encountered each type of crossing event twice, resulting in a total of 24 data points per event type.

\subsubsection{Vehicle Stopping}

\begin{figure}[h]
    \centering
    \begin{subfigure}{0.48\textwidth}
        \centering
        \includegraphics[width=\textwidth]{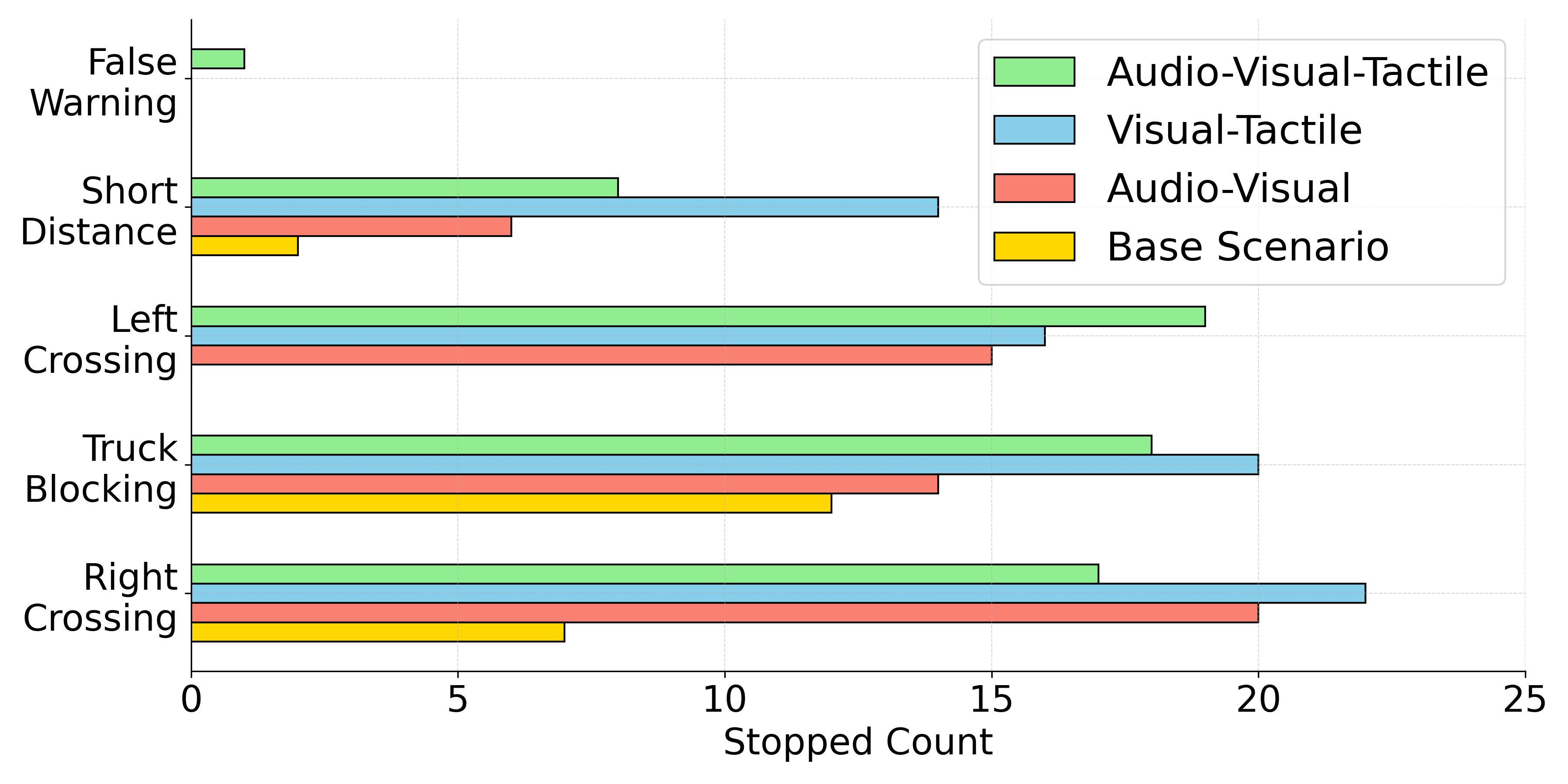}
        \caption{}
        \label{fig:stopping_count}
    \end{subfigure}
    \hfill
    \begin{subfigure}{0.48\textwidth}
        \centering
        \includegraphics[width=\textwidth]{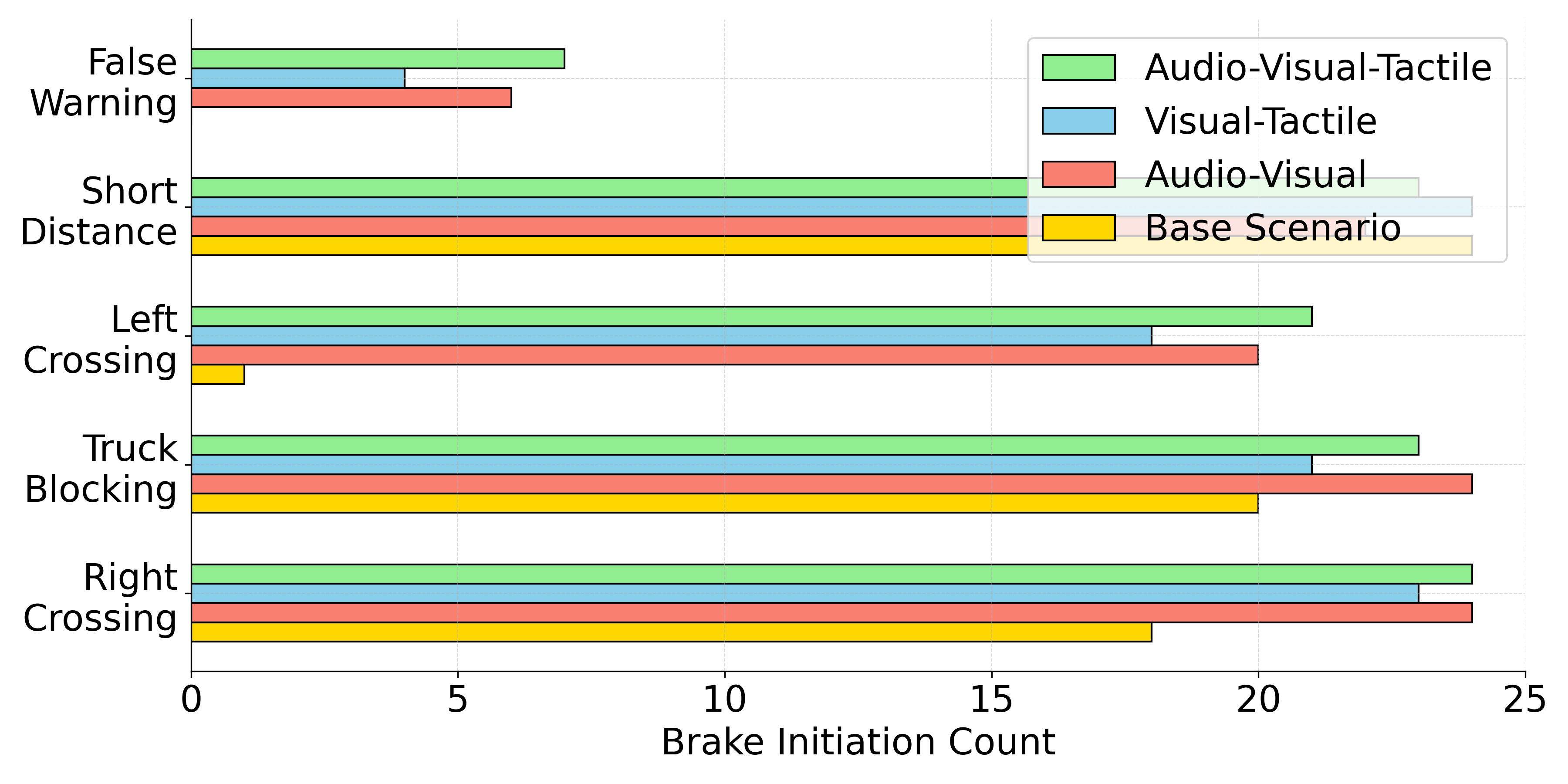}
        \caption{}
        \label{fig:brake_init}
    \end{subfigure}
    \caption{(a) Stopping vehicles count and (b) brake initiation count upon road events across scenarios}
    \label{fig:combined_figures}
\end{figure}

We first evaluated the impact of the alert system on vehicle stopping behavior (Figure~\ref{fig:stopping_count}). 

Our results indicate that the number of vehicles stopping upon receiving warnings significantly exceeded the number of stops in the no-warning condition across all crossing event types. In the base scenario without any warning, approximately half of the participants stopped and waited for pedestrians in the ``Right Crossing'' and ``Truck Blocking'' events. However, only two participants stopped in the ``Short Distance'' event. For the ``Left Crossing'' event, no participants stopped, likely due to pedestrians from their left side may have been outside of their immediate focus. 

In contrast, under conditions with alerts, around 80\% of participants stopped for the ``Left crossing'', ``Truck blocking'' and ``Right crossing'' events, while around one-third participants stopped for ``Short distance.'' 

Interestingly, we did not observe significant differences in stopping behavior between the three warning modalities. Among the four pedestrian jaywalking events, the ``VT'' modality resulted in the highest number of stops, except for the ``Left Crossing'' event. However, these differences were not statistically significant when compared to the other two modalities. It is also worth noting that participants may overreact to crossing alerts. One participant stopped for a false alarm during the session with ``AVT'' warning modality, whereas no participants with other warning modalities stopped. This suggests that combining all three warning modules may increase participants' caution and anxiety to the point where they may react to false alarms, which we further validate in our questionnaire described in Section~\ref{sec:attitudes}.



\subsubsection{Brake Initiation}
We examined the number of brake initiations upon receiving pedestrian crossing alerts or observing pedestrian jaywalking across driving sessions. 

In the ``Short Distance,'' ``Truck Blocking,'' and ``Right Crossing'' events, we observed similar numbers of participants applying brakes across both the warning scenarios and the baseline scenario without any warnings. However, in the baseline condition for the ``Left Crossing'' event, only one participant initiated braking. This aligns with our earlier observation that no participants stopped during the ``Left Crossing'' event in the no-warning condition. It suggests that only a small fraction of participants noticed the pedestrian jaywalking from the left or recognized it as a potential collision risk.

Conversely, when participants received alerts for the ``Left Crossing'' event, approximately 80\% applied the brakes and came to a stop. Such difference underscores the effectiveness of the alert system in drawing participants' attention to crossing pedestrians, particularly in scenarios where pedestrians may be less noticeable.

For the ``False Warning'' events where no pedestrians cross, no participant initiated brakes in baseline scenario. However, around 20\% participants applied brakes in driving sessions with warnings although only one stopped, indicating cautious response toward alerts.



\subsubsection{Average Stopping Distance}
\begin{figure}[h]
    \centering
    \begin{subfigure}{0.48\textwidth}
        \centering
        \includegraphics[width=\textwidth]{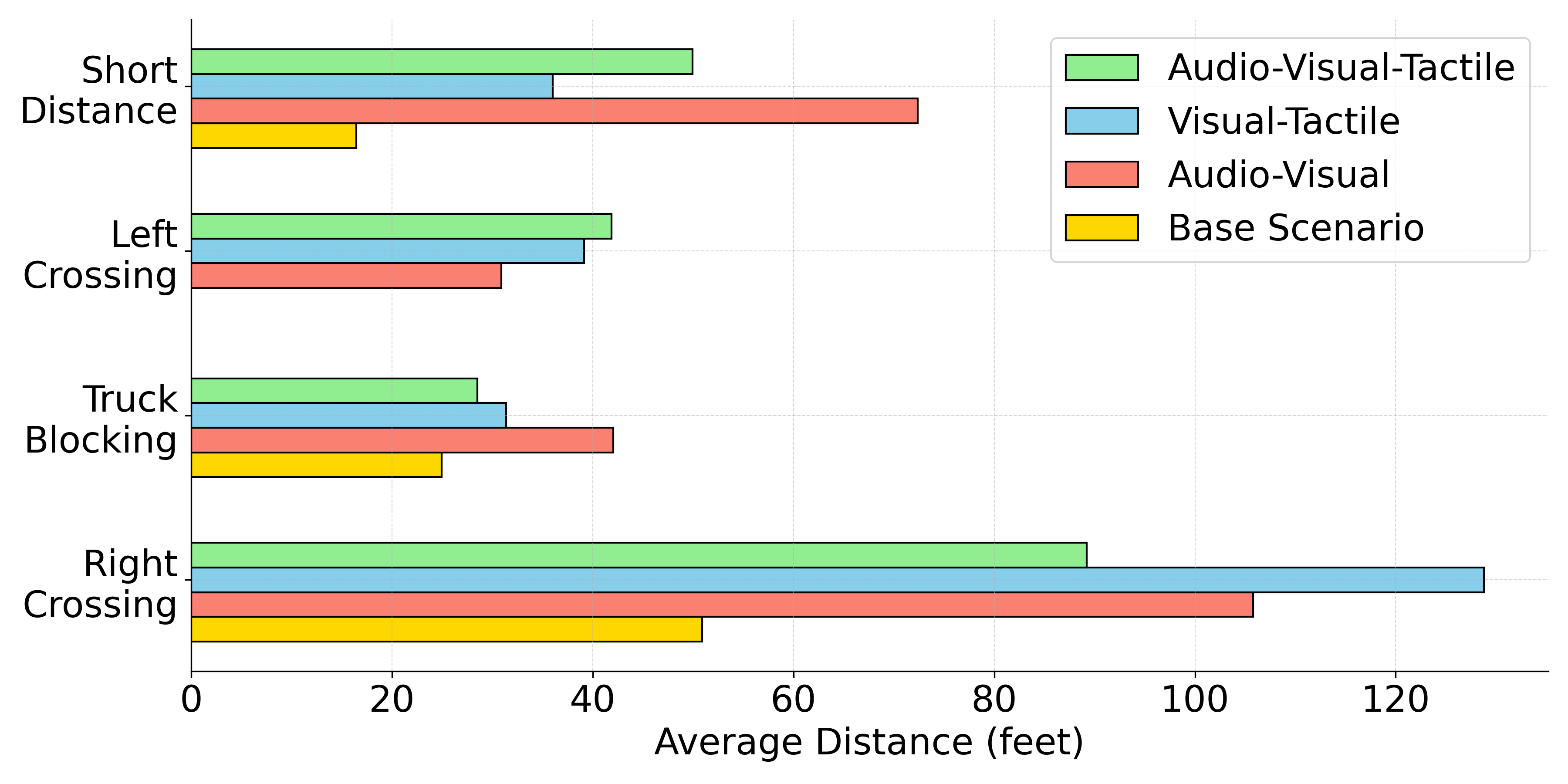}
        \caption{}
        \label{fig:average_distance}
    \end{subfigure}
    \hfill
    \begin{subfigure}{0.48\textwidth}
        \centering
        \includegraphics[width=\textwidth]{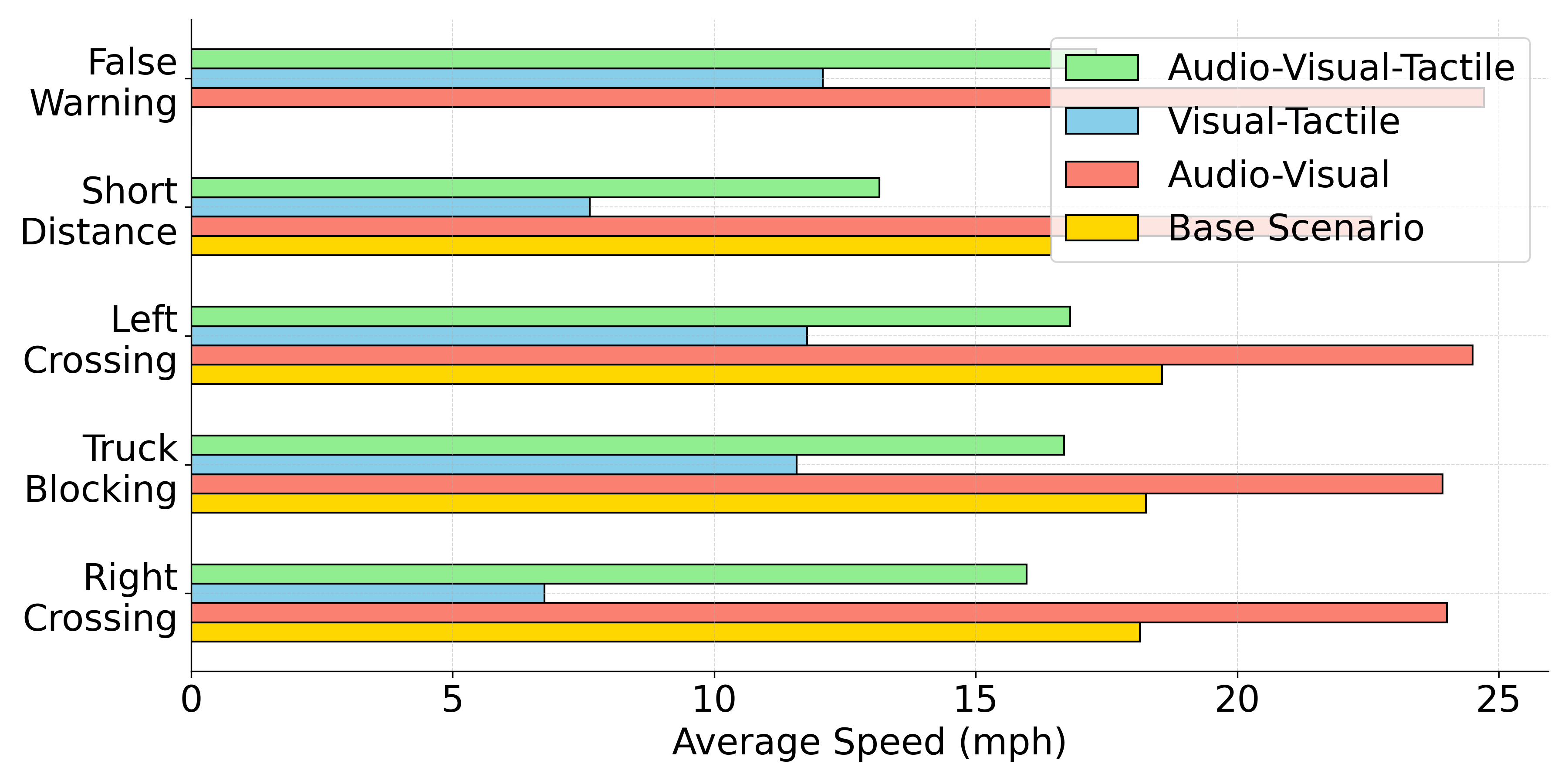}
        \caption{}
        \label{fig:average_speed}
    \end{subfigure}
    \caption{(a) Average distance for each event when car stops and (b) average speed after each event for each scenario.}
    \label{fig:combined_distance_speed}
\end{figure}

The average distance between the vehicle and the pedestrian at the point of vehicle stopping varied significantly across different scenarios (Figure~\ref{fig:average_distance}). For this analysis, we excluded the ``False Alarm'' events, focusing solely on pedestrian jaywalking.

Our findings indicate that all alert modalities significantly enhanced stopping distances compared to the baseline, with the AV modality resulting in the most substantial increase. The base scenario without any warning had the shortest average distance of 31.30 feet. In contrast, the warning modalities significantly increased the stopping distance, with the AV modality yielding the largest average distance at 59.89 feet, followed closely by the VT at 59.28 feet, and the AVT modality at 49.80 feet. 

Regardless of warning modalities, participants may apply brakes differently depending on the visibility of pedestrians. Among the event types, the ``Right Crossing'' events resulted in the longest stopping distances, while the ``Truck Blocking'' events produced the shortest, even shorter than those observed in the ``Short Distance'' events. Since participants received alerts at the same distance from pedestrians in all scenarios, this suggests that participants applied the brakes later or less forcefully in situations where they could not clearly see the pedestrians, such as during ``Truck Blocking'' events. Similarly, the ``Left Crossing'' event, where pedestrians were less obvious, also resulted in relatively shorter distances. In contrast, in the ``Right Crossing'' events, where the pedestrian was clearly visible, participants responded more quickly and stopped at a greater distance.


\subsubsection{Average Speed}
The average speed after one event completes and before the next event starts provides insights into how pedestrian crossing alerts impact driving behavior. 

As different warning modalities may influence participants' driving behavior in varying ways, the average speed after each event reflect these potential changes, such as how cautious drivers are in anticipation of future events. 
Our findings indicate that VT warning modality may have the strongest effect on increasing driver awareness of potential future collisions.
Participants who experienced the AV warning modality maintained the highest average speed, while those in the VT modality showed the lowest. In comparison, the base scenario, where no warning was provided, resulted in slightly higher average speeds than both the AVT and VT modalities. 


\subsubsection{Statistical Analysis}
We conducted an Analysis of Variance (ANOVA~\cite{nishishiba2014comparing}) test to evaluate the performance measures in the scenarios and use the base scenario as the control group. ANOVA allows for the comparison of means across multiple groups to identify statistically significant differences between them and the results are presented in Table~\ref{tab:comparison}. The null hypothesis ($H_0$) is that there is no significant difference in the means of performance measures among the scenarios. The results included F-statistics, which represent the ratio of the variance between groups and the variance within groups. The greater the F-statistic value, the more likely that the differences found are statistically significant. From the F-statistic, a p-value is computed, based on a statistical distribution known as the F-distribution under the condition that the null hypothesis is true. 

For Average Distance, the AV modality had the highest improvement compared to the base scenario, while the AVT and VT modalities also showed differences. For Stopped Count, the largest increase in stopped vehicles was observed in the VT modality followed by AVT and AV scenarios. The Enhancement is the difference between the one modality's average and the base average, with positive values representing improved performance. In the AV scenario, the enhancement for average distance was 31.92 feet, which means the vehicles stopped 31.92 feet farther from the pedestrian than in the base condition. A similar positive enhancement in the stopped count would suggest that more vehicles acted on the warning system in the test scenarios compared to the base scenario. For instance, the VT modality showed an enhancement of 11 vehicles stopped, indicating improved performance over the base condition.

\begin{table}[h]
    \centering
    \footnotesize
    \begin{tabular}{ c c c c c c c }
        \toprule
        \textbf{Scenario} & \textbf{Variable (units)}& \textbf{F-statistic} & \textbf{p-value} & \textbf{Base Scenario Average} & \textbf{Scenario Average} & \textbf{Enhancement}  \\
        \midrule
\rowcolor{myblue}
        AVT& Average Distance (ft)& 1.686& 0.251& 31.11& 52.91& 22.8\\
        VT& Average Distance (ft)& 1.206& 0.333& 30.11& 58.84& 28.73\\
\rowcolor{myblue}
       AV& Average Distance (ft)& 2.634& 0.169& 30.11& 62.03& 31.92\\
        VT& average Stopped Count (number)& 5.1& 0.061& 7& 18& 11\\
\rowcolor{myblue}
        AVT& average Stopped Count (number)& 3.986& 0.086& 7& 15.75& 8.75\\
        AV& Stopped Count (number)& 2.563& 0.154& 7& 13.75& 6.75\\
        \bottomrule
    \end{tabular}
    \caption{Comparison of Scenarios with the Base Scenario}
    \label{tab:comparison}
\end{table}


\subsection{Participant Feedback in Post-Session Questionnaire}

\subsubsection{Satisfaction toward Pedestrian Alert System}
\begin{figure}[h]
    \centering
    \begin{minipage}[b]{0.48\textwidth}
        \centering
        \includegraphics[width=\textwidth]{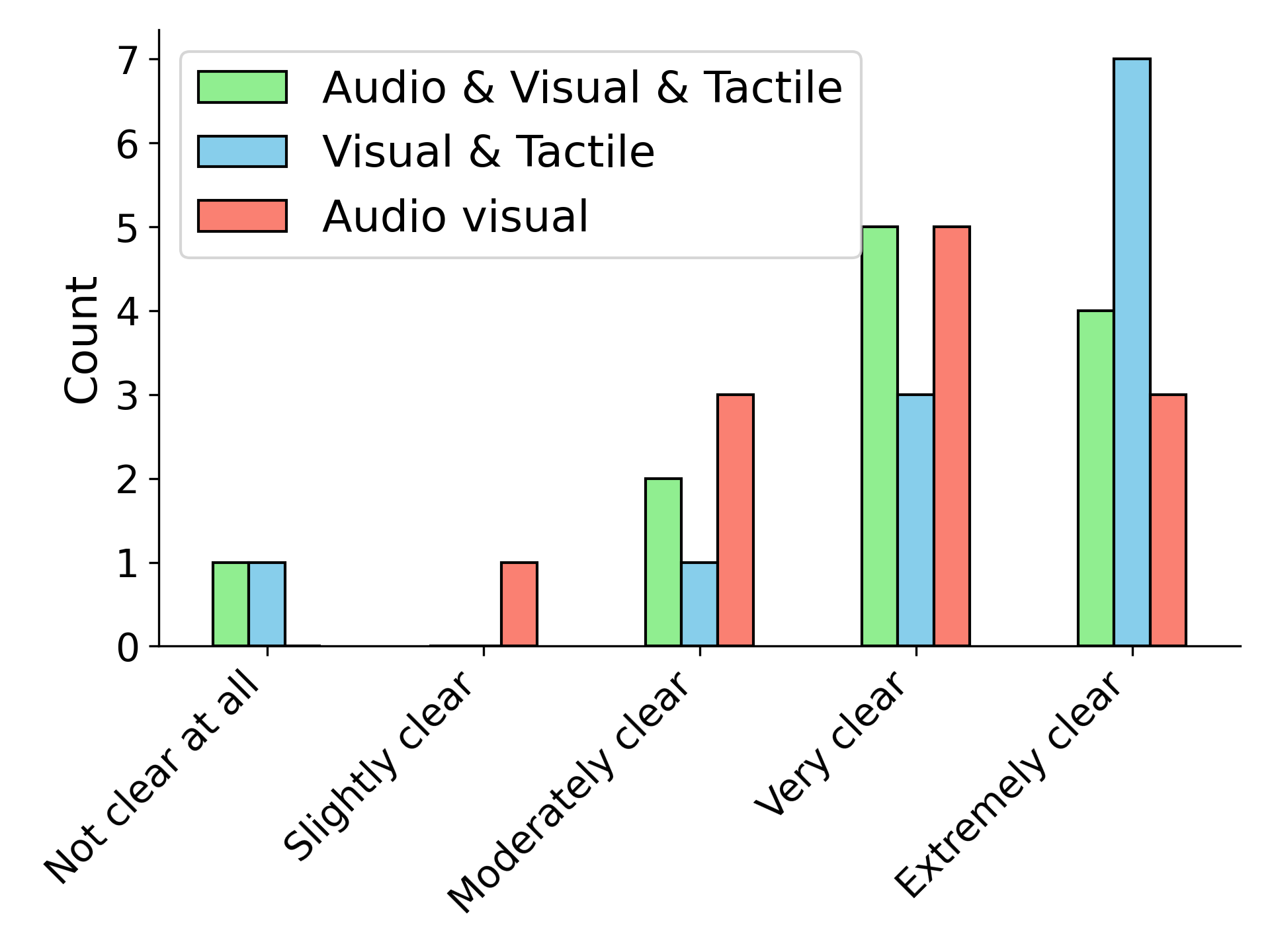}
        \caption{Q: How clear were the pedestrian crossing alerts?}
        \label{fig:clarity}
    \end{minipage}
    \hfill
    \begin{minipage}[b]{0.48\textwidth}
        \centering
        \includegraphics[width=\textwidth]{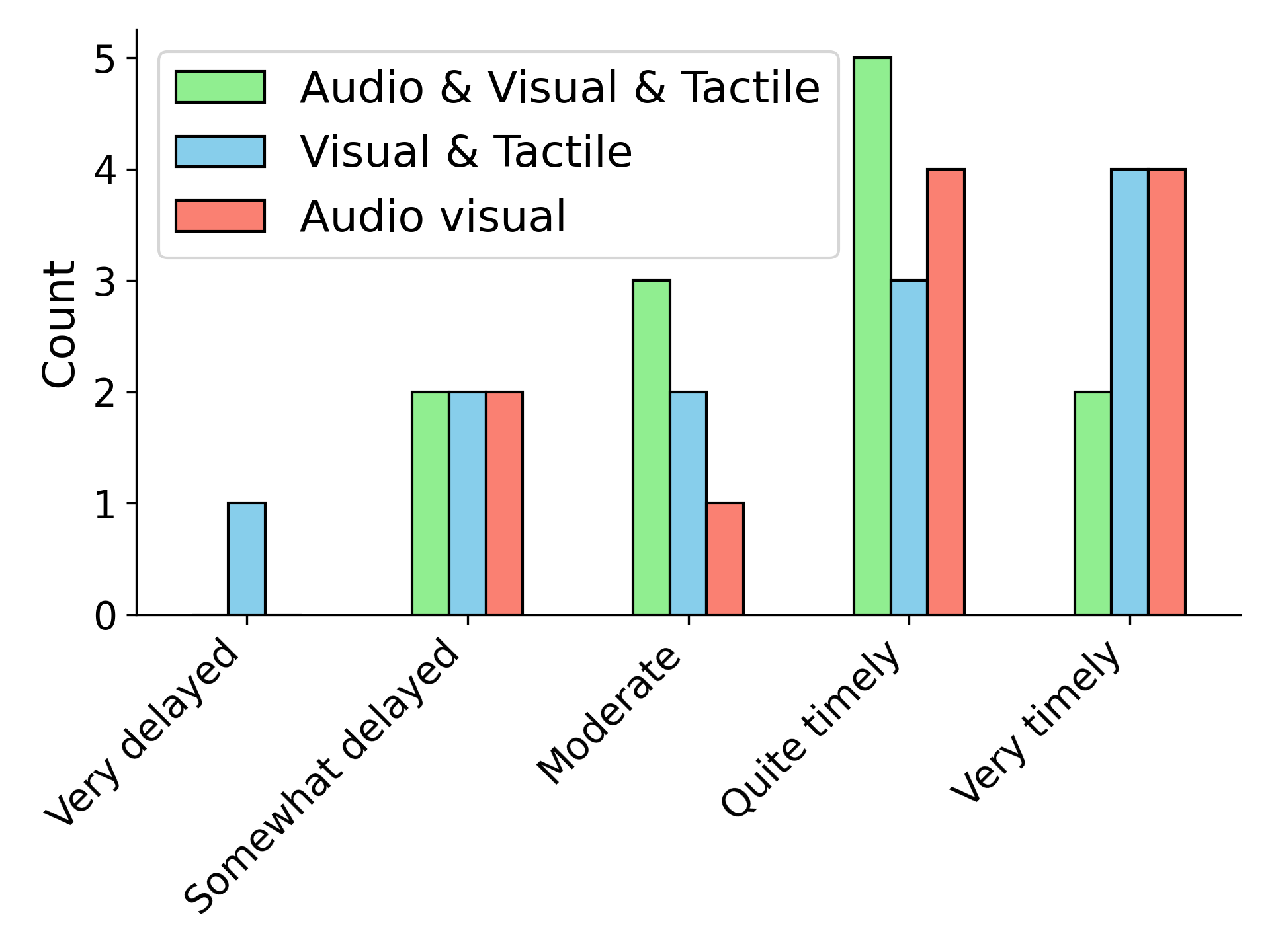}
        \caption{Q: How timely were the the pedestrian crossing alerts?}
        \label{fig:timeliness}
    \end{minipage}
    
    \vspace{0.5cm} 
    
    \begin{minipage}[b]{0.48\textwidth}
        \centering
        \includegraphics[width=\textwidth]{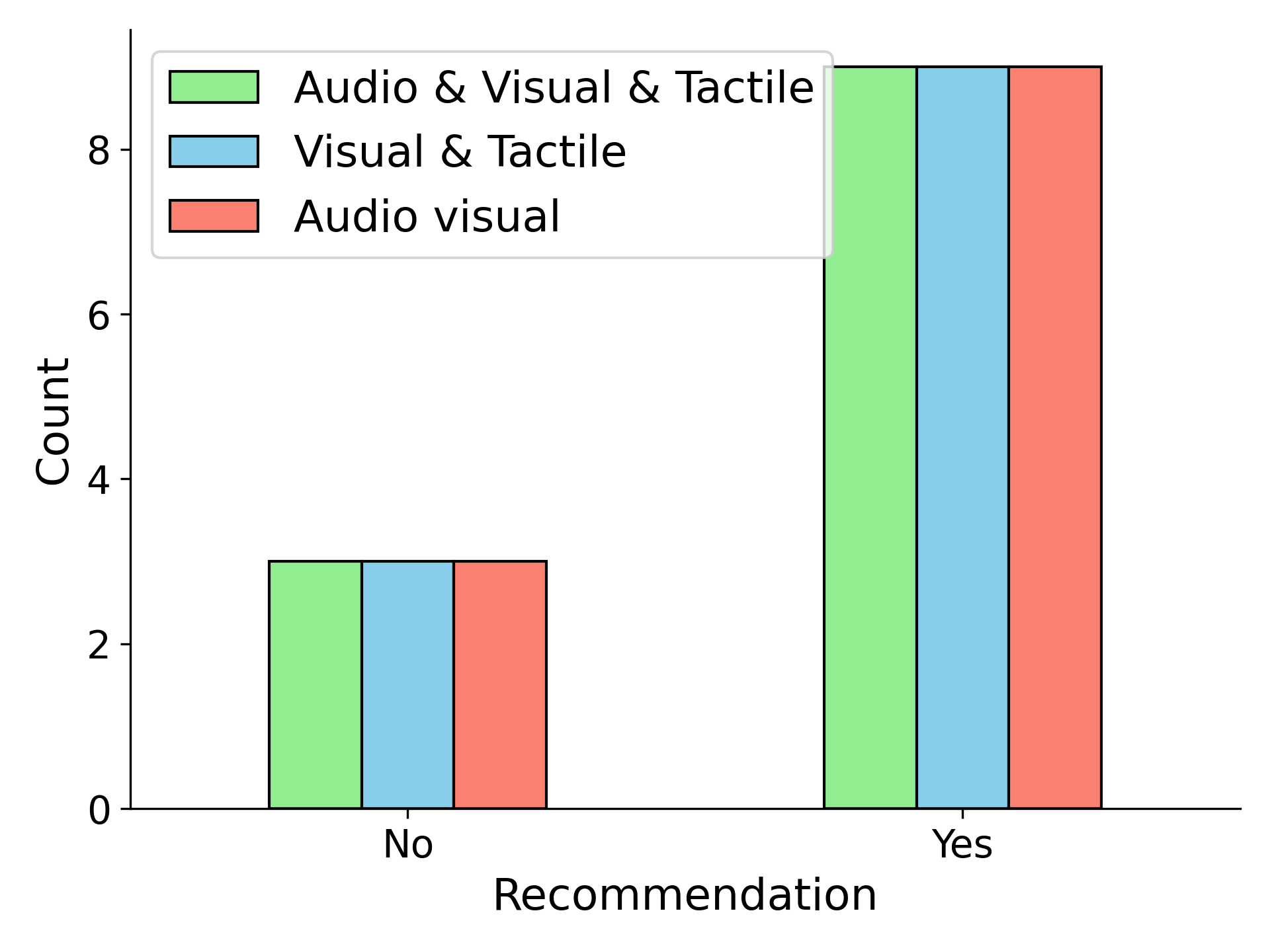}
        \caption{Q: Would you recommend this pedestrian warning system to other drivers? }
        \label{fig:recommendation}
    \end{minipage}
    \hfill
    \begin{minipage}[b]{0.48\textwidth}
        \centering
        \includegraphics[width=\textwidth]{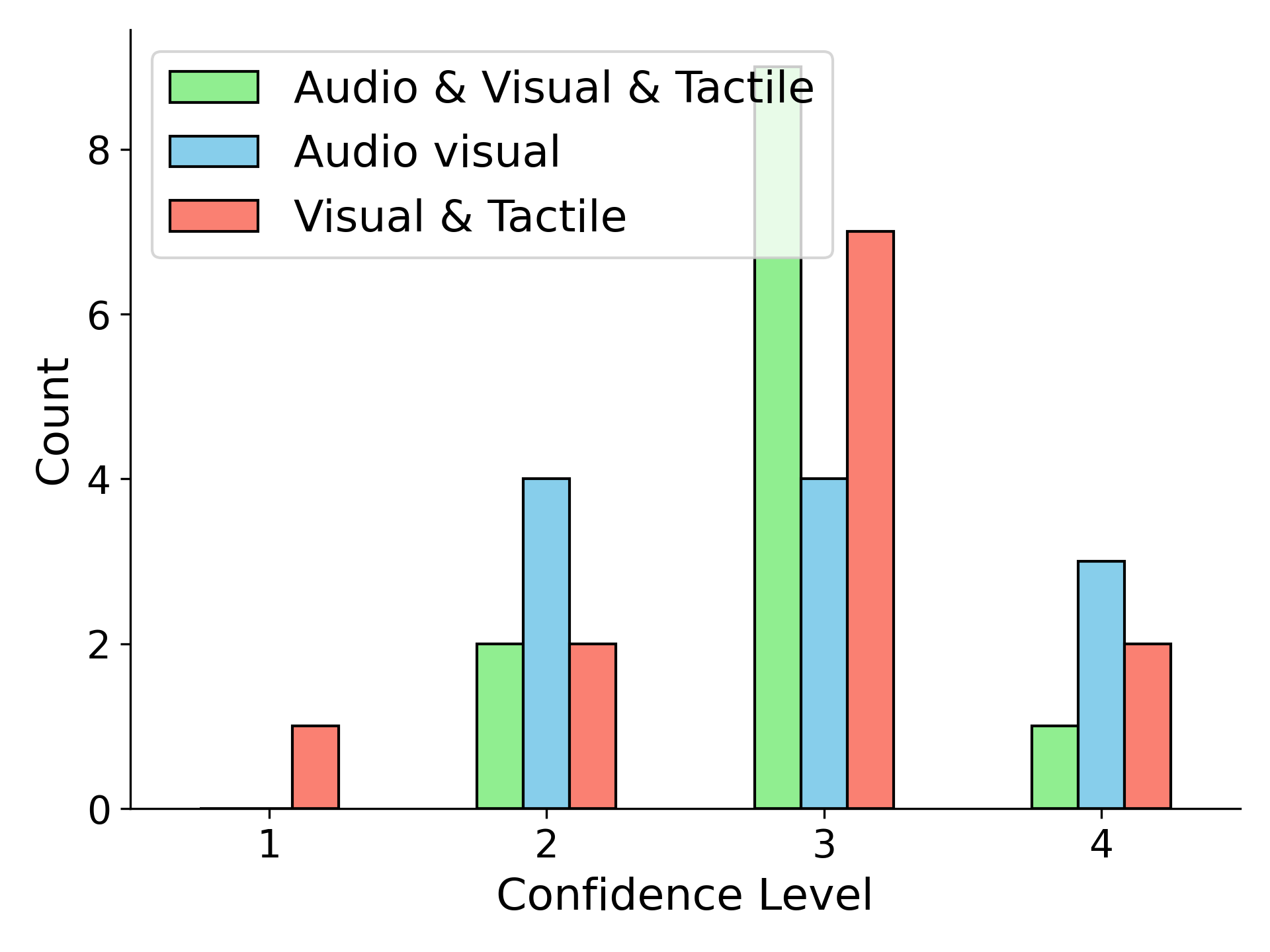}
        \caption{Q: How would you rate your overall confidence in the system for pedestrian crossing alerts? (4 is the highest)}
        \label{fig:confidence}
    \end{minipage}
\end{figure}

We evaluated participants' satisfaction towards the pedestrian alert system through multiple-choice questions in the post-session questionnaire. We evaluated several key aspects, including the clarity, timeliness, and overall confidence in the system, as well as whether they would recommend the system to others. Below is a summary of the key findings from the post-session questionnaire:
\begin{itemize}

\item \textbf{Clarity of Alerts:}
As Figure~\ref{fig:clarity} shows, participants highly rated the clarity of the system alerts for pedestrian crossings across all scenarios. While the majority of participants in the AVT scenario found the alert to be ``Extremely clear,'' many in the VT and AV scenarios participants considered the alert either ``Very clear'' or ``Extremely clear.''

\item \textbf{Timeliness of Alerts:}
Figure~\ref{fig:timeliness} shows the timeliness of pedestrian crossing alerts. Of all scenarios, the majority rated the alerts either ``Quite timely'' or ``Very timely.'' It is also worth mentioning that, out of all scenarios assessed, the highest percentage rating for ``Quite timely'' is the AVT scenario.

\item \textbf{System Recommendation:}
When asked whether they would suggest the pedestrian alert system to other drivers, 75\% of participants in every scenario responded positively.

\item \textbf{Overall Confidence:}
In general, confidence in the pedestrian warning systems was on a scale of 1 to 4, where 4 was also mapped to ``Very Confident.'' The confidence levels were generally high for all scenarios, while the AVT scenario received the highest level of confidence, followed by the VT scenario. 
\end{itemize}

\subsubsection{Perceptions toward False Alarms}\label{sec:attitudes}
We analyzed participants' perceptions towards false alarms during driving sessions through an open-ended question in the post-session questionnaire. Participants were asked one open-ended question: ``How did the system's \textbf{false alerts} or \textbf{failures to detect} pedestrian crossing scenarios affect your driving?  (you do not need to answer if you have no alert)''. The purpose of this question was to assess how participants recognized and reacted to false alarms, given that they were not informed beforehand about the pedestrian alert system or the possibility of false alarms.

To ensure a comprehensive analysis, two researchers reviewed the participants' responses and identified eight common perceptions. Then each researcher independently coded each response using the perceptions as a basis. 
We show the results in Table~\ref{tab:attitudes_results} and describe each perception as follows: 

\begin{itemize}
    \item \textbf{Confusion or Uncertainty:} 13 participants expressed confusion regarding the alerts' meaning, with 10 participants unsure if they had missed a pedestrian and 3 others suspecting the system was signaling something unrelated, such as speeding.
    \item \textbf{Stress and Anxiety:} 3 participants who experienced multi-modal alert (including visual, auditory, and haptic cues) reported heightened stress and anxiety while driving. Participant 23 said: ``It made me extremely anxious while driving that I was going to hit something.'' False alarms increased their stress levels, while alerts with multiple modalities amplify their feelings of anxiety. 
    \item \textbf{Increased Caution:} 18 Participants indicated that the false alarms caused them to drive more cautiously. They mentioned checking their surroundings more frequently, slowed down, or lifted off the accelerator in response to these alerts. 
    \item \textbf{Distracted:} 8 participants reported being distracted by the false alarms, particularly when they were unable to immediately identify the cause. This distraction occasionally shifted their focus away from driving, introducing potential safety risks.
    \item \textbf{Immediate Reactions to Brake:} 6 participants reported that they instinctively applied the brake upon receiving an alert, even if they did not see a pedestrian nearby. 
    \item \textbf{Reduced Trust:} Seven participants mentioned that repeated false alerts reduced their trust in the system. Inconsistent or inaccurate alerts led them to question the system’s reliability and accuracy, making them more skeptical of its usefulness in real driving scenarios. 
    \item \textbf{Alerting Fatigue/Ignoring Alerts:} As the number of false alarms accumulated, 3 participants reported becoming desensitized to the alerts. This led them to ignore subsequent warnings, including potentially legitimate ones, commonly referred to as ``alert fatigue.''
    \item \textbf{Momentary Startle:} 3 participants reported feeling momentarily startled when the alerts first sounded, though they quickly recovered and resumed driving.
\end{itemize}

\begin{table}[ht]
\centering
\resizebox{\linewidth}{!}{%
\begin{tabular}{lcccccccc}
\toprule
\textbf{Modality} & \textbf{Confusion} & \textbf{Stress/} & \textbf{Increased} & \textbf{Distracted} & \textbf{Immediate} & \textbf{Reduced} & \textbf{Alert} & \textbf{Momentary} \\
& & \textbf{Anxiety} & \textbf{Caution} & & \textbf{Brake} & \textbf{Trust} & \textbf{Fatigue} & \textbf{Startle} \\
\midrule
\rowcolor{myblue}
\textbf{A+V} & 4 & 0 & 6 & 3 & 1 & 1 & 1 & 2 \\
\textbf{V+T} & 6 & 0 & 6 & 2 & 2 & 3 & 2 & 1 \\
\rowcolor{myblue}
\textbf{A+V+T} & 3 & 3 & 6 & 3 & 3 & 3 & 0 & 0 \\
\bottomrule
\end{tabular}
} 
\caption{36 Participants' Perceptions toward False Alarms across Modality. Abbreviations: A+V = Audio \& Visual, V+T = Visual \& Tactile, A+V+T = Audio, Visual \& Tactile.}
\label{tab:attitudes_results}
\end{table}




\section{Conclusions, Limitations, and Recommendations}
The driving simulator study evaluated the impact of different pedestrian crossing alert modalities on driver behavior. The findings suggest that all warning modalities (AV, VT, and AVT) enhance driver response compared to the base condition of no warning when the distance to pedestrians and stop count is considered. The enhancement in the average stopping distance ranged from +75\% to +105\% across the scenarios, and the enhancement in stopped count ranged from +96\% to +157\%. The VT warning showed the most substantial increase in the stopping count with an enhancement of 11 vehicles over the base condition. For stopping distance, AV warning showed the most improvement with an additional 31 feet over the base condition. Eighteen participants across modalities indicated that they drove more cautiously at times slowing down or checking surroundings when it was not necessary. Seven participants indicated that repeated false alarms reduced their trust in the system, leading to alert fatigue and, in some cases, ignoring subsequent warnings altogether. Participants who experienced the AVT system also reported the highest levels of stress and distraction, suggesting that the intensity of multi-sensory alerts may contribute to heightened anxiety. 

While the study provides valuable insights, it also has some limitations. One challenge was motion sickness, which made 19 out of 67 participants we recruited unable to complete the driving session. Additionally, the semi-controlled environment of the simulator may have influenced participants to be more cautious than they would be in real-world scenarios, potentially reducing risk-taking behaviors. Future research efforts with a larger number of participants are recommended to confirm the findings of this research.


\bibliographystyle{ACM-Reference-Format}
\bibliography{human_factor}

\end{document}